\documentclass[
 reprint,
 amsmath,amssymb,
 aps,
 floatfix,
 pre,
 superscriptaddress
 ]{revtex4-2}

\usepackage{microtype}
\usepackage{bm}
\usepackage{physics}
\usepackage{mathtools}
\usepackage{nicefrac}
\DeclareMathOperator\arctanh{arctanh}
\usepackage{graphicx}
\usepackage{subfigure}
\usepackage[sort&compress]{natbib}
\usepackage[hidelinks]{hyperref}
\usepackage{cleveref}

\begin{document}

\title{Stochastically bistable growth and decay in the Togashi-Kaneko model}

\author{Jeremy R. Worsfold}
\email{j.worsfold@unsw.edu.au}
\affiliation{
    School of Physics, UNSW, Sydney, NSW 2052, Australia.
}
\affiliation{
    EMBL Australia Node in Single Molecule Science, School of Biomedical Sciences, UNSW, Sydney, NSW 2052, Australia. 
}

\author{Richard G. Morris}
\affiliation{
    School of Physics, UNSW, Sydney, NSW 2052, Australia.
}
\affiliation{
    EMBL Australia Node in Single Molecule Science, School of Biomedical Sciences, UNSW, Sydney, NSW 2052, Australia. 
}
\affiliation{
    ARC Centre of Excellence for the Mathematical Analysis of Cellular Systems, UNSW Node, Sydney, NSW 2052, Australia. 
}

\date{\today}

\begin{abstract}
The two-state Togashi-Kaneko model demonstrates how, at finite system sizes, autocatalysis can lead to noise-induced bistability between the cellular concentrations of different molecular species.
Here, we show that, in the biologically relevant scenario of species-dependent export rates, the nascent stochastic switching between molecular species also drives a concomitant switching between periods of growth or decay in the total population size.
We demonstrate this behavior using stochastic simulations as well as the numerical integration of a Fokker-Planck equation that approximates the finite system-size limit.
By combining piecewise-deterministic-Markov and linear-noise approximations, we further find analytic expressions for the stationary distributions of the different molecular species when stochastic switching is faster than the dynamics in the total population size.
We envisage that other models in the voter class--- including spin systems, flocking and opinion dynamics--- may also exhibit aperiodic growth and decay in population size, as well as be amenable to similar techniques.

\end{abstract}

\maketitle

\section{Introduction}

In the last two decades, it has become increasingly evident that proper consideration of intrinsic noise is integral for a complete understanding of many biochemical, ecological, and sociological systems \cite{horsthemke2006noise,biancalani_noiseinduced_2014,jhawar_noiseinduced_2020,herrerias-azcue_consensus_2019, constable_demographic_2016,sardanyes_noiseinduced_2018}. Of particular interest is the phenomenon of noise-induced bistability, which has been observed in the collective behavior of ants \cite{biancalani_noiseinduced_2014} and fish \cite{jhawar_noiseinduced_2020}, and has been hypothesized to drive consensus in opinion dynamics \cite{morris_growthinduced_2014, herrerias-azcue_consensus_2019}. 
In the context of molecular biology, noise-induced bistability was popularized by Togashi and Kaneko (TK), and their eponymous model of autocatalytic chemical reactions in a cell \cite{togashi_transitions_2001, biancalani_noiseinduced_2012}.

Noise-induced bistability arises from intrinsic, finite-size fluctuations that are state-dependent, leading to bimodal probability distributions whose extrema do not coincide with the fixed points of the system's deterministic dynamics. This gives rise to a stochastic switching between the states localized at either of these extrema, mimicking conventional bistability in systems that have two deterministic fixed points and additive noise. (In one dimension the latter correspondence can be made exact).

In this context, we revisit the TK model and, in particular, species-dependent fluxes. 
These aim to reflect the biological reality that rates transport, synthesis and /or degradation are, generally speaking, dependent on the specific molecule, or protein at hand (see \textit{e.g.} \cite{pakdel_exploring_2018,cross_delivering_2009,groll_molecular_2005,dimou_unconventional_2018}).
Notably, however, whilst the subject of species-dependent import has been studied in the TK model \cite{bibbona_stationary_2020, gallinger_asymmetric_2024}, the subject of species-dependent degradation and/or export has remained open. 

A potential reason for this, we argue, is that such a parameter choice acts to break the mold of most existing models of noise-induced bistability. These models broadly fall into one of three possible scenarios, characterized in terms of the dynamics of the total population size, $N$. 
In the first, and simplest, case, the system is closed, so $N$ is a constant \cite{biancalani_noiseinduced_2014,jhawar_noiseinduced_2020,herrerias-azcue_consensus_2019,houchmandzadeh_exact_2015a,jafarpour_noiseinduced_2015,jafarpour_noiseinduced_2017}. 
In the second case, $N$ undergoes trivial, Gaussian fluctuations \cite{biancalani_noiseinduced_2012, bibbona_stationary_2020, gallinger_asymmetric_2024, togashi_transitions_2001,saito_theoretical_2015}.
Finally, in the third case, $N$ increases in time according to some growth protocol  \cite{morris_growthinduced_2014,cremer_evolutionary_2011,crosato_dynamical_2023,melbinger_evolutionary_2010}. In all three cases, the model setup typically ensures that the dynamics of $N$ is independent of the noise-induced switching. 

By contrast, taking the export rates of the TK model to be species-dependent \emph{couples} the rate of change of $N$ to the relative proportion of each species. 
As a result, we observe that the usual bistable switching between two metastable points is replaced with two alternating metastable flows of growth and decay in the system size.
The resulting non-equilibrium stationary state (NESS) has a non-zero, circular probability current which causes trajectories to perform stochastic loops in phase-space.
Whilst these trajectories have no well-defined period, since the bistable switching has exponentially-distributed waiting times, their loops are reminiscent of other stochastic systems with circular currents, such as stochastic limit cycles and cycles due to stochastic amplification \cite{butler_fluctuationdriven_2011,biancalani_stochastic_2010}. 

The remainder of this article is organized as follows.
First, after setting up the model and fixing notation, we use a Kramers-Moyal expansion of the Master equation to obtain a Fokker-Planck equation and its corresponding coupled stochastic differential equations (SDEs) \cite{gardiner_handbook_1985}.
These capture the stochastic evolution of the system at finite system sizes.
By using Gillespie simulations \cite{gillespie_exact_1977a} as well as finite-element numerical integration of an approximate Fokker-Planck equation \cite{verdugo_software_2022}, we characterize the central phenomena of our article: species-dependent export rates couple bistability in species with flows of population growth and/or decay. 
Notably, we find that, when autocatalytic reactions are faster than molecular import or export, the changing metastable flows in total population can be approximated as a Piecewise Deterministic Markov Process (PDMP) \cite{faggionato_nonequilibrium_2009}, with fast switches in the relative proportion controlling the direction of the metastable flow.
Combining this insight with with recent developments in the study of PDMPs for finite populations \cite{faggionato_nonequilibrium_2009,hufton_intrinsic_2016,hufton_model_2019}, we are able to solve for the stationary distribution in this fast-switching regime.
We then incorporate asymmetric import into this PDMP approximation.
In this PDMP limit, the stationary expectation of the relative proportion of each species, $y$, is shown to differ from the deterministic limit, a result not seen in typical noise-induced bistable systems. 
We find strong agreement with our stochastic simulations up to a point where the expected value of $y$ transitions to the deterministic limit.

\section{Generalized two-state Togashi-Kaneko model}

\begin{figure}
    \centering
    \includegraphics[width=8.6cm]{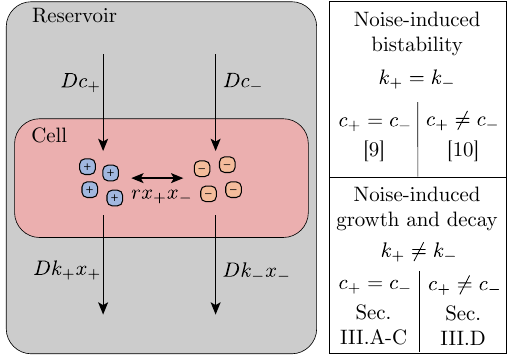}
    \caption{\textbf{Two-state Togashi-Kaneko model (TK2)}. 
    Left: Generic representation of the import, autocatalysis, and export reactions that make up the TK2 model. 
    Right: Existing literature has so far focused on the effects intrinsic noise and the resulting noise-induced bistability of molecular concentrations, either with symmetric or asymmetric import (top). Here, we focus on the biologically-relevant case of export rates that depend on the type of molecule (\textit{e.g.}, proteins) where we show that noise-induced bistability becomes coupled to population growth and decay (bottom). 
    }
    \label{fig:model}
\end{figure}

The general Togashi Kaneko model considers a hypothetical cell in which a number of molecular species interact through autocatalytic reactions.
We consider the two-species case--- so-called `TK2'--- denoting molecules of each species by $X_\pm$.
There are three broad components to the model.
\begin{enumerate}
    \item First, import from the external environment. This is assumed to be an infinite reservoir, such that the external concentrations of each species, $c_\pm$, are held constant.
    \item Second, once inside the cell, molecules undergo autocatalytic reactions of the form $2X_+ \leftarrow X_++X_-\rightarrow 2X_-$. The rate constant, $r$, of these reactions is taken to be symmetric for simplicity. (Small asymmetry has been considered elsewhere \cite{gallinger_asymmetric_2024}).
    \item Third, the molecules are either transported back to the reservoir or they are degraded, a process we generically refer-to as export. The purpose of this article is to characterize the ramifications when such export is species dependent, since a biological cell will generally degrade and/or transport different types of molecules at distinct rates.
    To capture this, we therefore write the export rate for species $X_\pm$ as $Dk_\pm$, where, following the literature, $D$ is referred-to as the diffusion constant (but could ultimately represent the rate constant for myriad biological mechanisms of transport and/or degradation).
\end{enumerate}
These processes, and their respective rates, are summarized in \cref{fig:model}.

When the cell volume $V$ is sufficiently large, the concentrations $x_\pm=n_\pm/V$ (where $n_\pm$ are the molecular counts of each species and $N=n_++n_-$ is the total population) are governed by a set of equations arising from deterministic mass-action kinetics 
\begin{align}
    \dot{x}_\pm = D(c_\pm -k_\pm x_\pm).
    \label{eq:mass_action}
\end{align} 
This two-dimensional dynamical system contains a single stable node at $x_\pm^*=c_\pm/k_\pm$.

\begin{figure*}
    \centering
    \includegraphics[width=17.2cm]{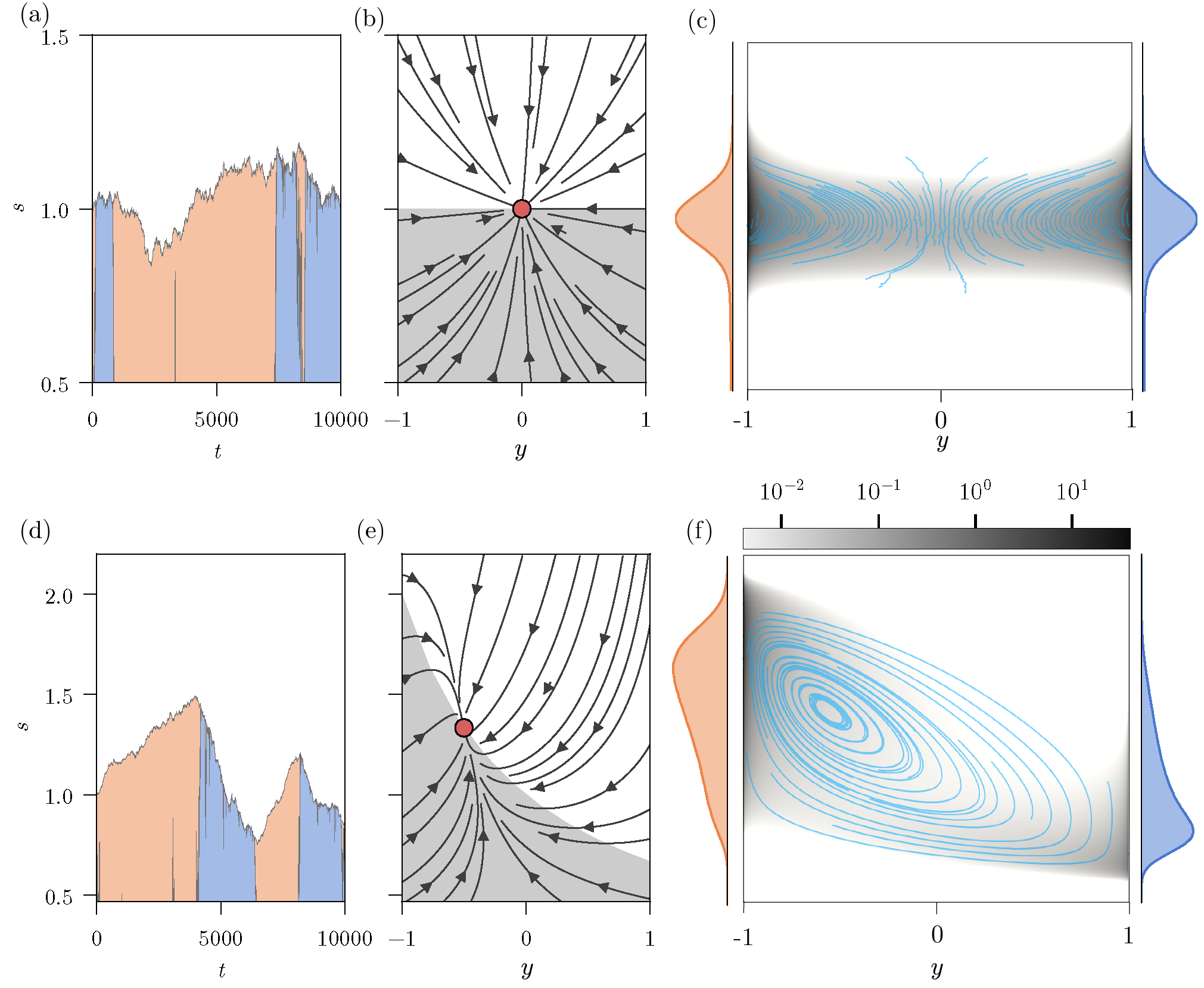}
    \caption{\textbf{Species-dependent export couples noise-induced bistability to population size}. (a,d): Gillespie simulations showing populations $x_+$ (blue) and $x_-$ (orange)  \textit{vertically stacked} with $\lambda=DV/2=0.04$. (a-c): Stochastic switching between dominant species due to noise-induced bistability when the export rates are identical ($\kappa=0$). (d-f): Stochastic switching between dominant species coupled to growth and decay when export rates are asymmetric ($\kappa=0.5$). (b,e): Vector field lines from the deterministic mass-action kinetic prediction with the red dot located at the fixed point while the gray and white regions indicate increasing and decreasing $s$, respectively. (c,f): Stationary probability distribution (grayscale) and streamlines in the current $\bm{J}$ (blue) of the Fokker Planck equation \cref{eq:currentJ} using the finite element method for $\lambda=0.5$ with the marginal distributions of $\int_{0}^1P(s,y)\dd y$ (right, blue) and  $\int_{-1}^0P(s,y)\dd y$ (left, orange).}
    \label{fig:vecflows}
\end{figure*}

However, as has been noted in \cite{biancalani_noiseinduced_2012, biancalani_noiseinduced_2014}, the dynamics encoded by (\ref{eq:mass_action}) are \emph{not} representative of the broader behavior of the TK model at finite volumes when the autocatalytic reactions are fast compared with the diffusion $D$.
To illuminate the role of intrinsic noise, and explore the consequences of species dependent export rates in this setting, our staring point is the governing Master equation.
Specifically, by writing $Q(\bm{x}|\bm{x}')$ as the transition rate from state $\bm{x}'$ to $\bm{x}$ with $\bm{x}=(x_+,x_-)^T$, we express the dynamics of the probability $P(\bm{x},t)$ as:
\begin{align*}
    \partial_tP(\bm{x},t) = \sum_{\bm{x}'\neq \bm{x}}\left[Q(\bm{x}|\bm{x}')P(\bm{x}',t)-Q(\bm{x}'|\bm{x})P(\bm{x},t)\right],
\end{align*}
where the possible transitions are given by
\begin{equation}
\begin{aligned}
    Q\left(x_+ \pm\frac{1}{V},x_- \mp\frac{1}{V}\bigg|x_+,x_-\right) & = r x_+ x_-, \\
    Q\left(x_\pm+\frac{1}{V}\bigg|x_\pm\right) & = Dc_\pm, \\
    Q\left(x_\pm-\frac{1}{V}\bigg|x_\pm\right) & = Dk_\pm x_\pm.
\end{aligned}
\end{equation}
Following \cite{biancalani_noiseinduced_2012}, we then perform a Kramers-Moyal expansion in  $1/V$. 
Truncating this expansion at second order and rescaling time $t/V\to t$, we obtain a Fokker Planck equation for the evolution of the species populations which is statistically equivalent to the following It\^o SDEs
\begin{align}
    \dot{x}_\pm = D(c_\pm - k_\pm x_\pm) + \frac{1}{\sqrt{V}}\eta_\pm(t), \label{eq:xSDE}
\end{align}
where $\eta_\pm(t)$ are Gaussian white noises with correlation given by
\begin{align*}
   \langle\eta_\pm(t)\eta_{\pm}(t')\rangle & = \left[D\left(c_\pm + k_\pm x_\pm\right) + 2rx_+x_-\right]\delta(t-t'), \\
   \langle\eta_\pm(t)\eta_{\mp}(t')\rangle & = -2rx_+x_-\delta(t-t').
\end{align*}

At this stage, and in anticipation of results to come, it is helpful to make three simplifying choices. First, and without loss of generality, we rescale the diffusion constant $D/r\to D$, noting only that $D$ is now dimensionless. Second, we change variables to the total concentration, $s=x_++x_-$, and the complementary relative proportion, $y=(x_+-x_-)/s$. Third, we explicitly parameterize the potential species dependence of import and export rates, writing $c_\pm=(1\pm\beta)/2$ and $k_\pm=1\pm\kappa$, respectively. (Note that the former implies that the total external concentration is set to $c_++c_-=1$).

Using It\^o's lemma to perform the change of variables, and again rescaling time--- now according to $Dt\to t$--- we obtain two SDEs of the form
\begin{subequations}
    \begin{align}
        \dot{s} &= A_1(s,y) + \eta_1(t),\label{eq:sSDE}\\
        \dot{y} &=  A_2(s,y) + \eta_2(t) \label{eq:ySDE}
    \end{align}
\end{subequations}
where the correlation in the white noise is given by  $\langle\eta_{i}(t)\eta_j(t')\rangle=\delta(t-t')\mathsf{B}_{ij}(s,y)$ and $\bm{A}=(A_1,A_2)^T$ with
\begingroup
\setlength\arraycolsep{10pt}
\begin{widetext}
\begin{equation}
\begin{aligned}
    \bm{A}(s,y) &= \begin{pmatrix}
        1 - s(1+\kappa y) \\
        \dfrac{\strut\beta- y}{s} - \kappa(1-y^2)
    \end{pmatrix} + \frac{1}{V} \begin{pmatrix}
        0 \\ \dfrac{\strut y-\beta-\kappa s(1-y^2)}{\strut s^2}
    \end{pmatrix}, \\
    \mathsf{B}(s,y) &= \frac{1}{V}\begin{pmatrix}
        1+s(1+\kappa y) & \kappa(1-y^2) + \dfrac{\strut\beta- y}{\strut s} \\
        \kappa(1-y^2) + \dfrac{\strut\beta- y}{\strut s} & \dfrac{\strut s(1-\kappa y)(1-y^2) + 1+y^2-2\beta y}{s^2}
    \end{pmatrix} + \frac{2}{DV}\begin{pmatrix}
        0 & 0 \\
        0 & 1-y^2
    \end{pmatrix}.
    \label{eq:AandB}
\end{aligned}
\end{equation}
\end{widetext}
\endgroup

If we take $V\to\infty$ while keeping $D$ constant, only the first term of $\bm{A}$ in \eqref{eq:AandB} remains and we recover the dynamical system predicted by mass action kinetics.
The corresponding single stable node is given by $(s^*,y^*)$, where
\begin{align}
     s^*= \frac{1-\beta  \kappa}{1-\kappa ^2},\quad y^* = \frac{\beta-\kappa }{1-\beta  \kappa}.\label{eq:deterministicmean}
\end{align}
In systems such as the TK model, however, this deterministic fixed point is not representative of the behavior as the noise strength increases.

\subsection{Noise-induced bistability}
\label{sec:NIbistability}

To describe the basis of noise-induced effects and to provide context for our results, we first revisit the case where $\kappa=\beta=0$--- \textit{i.e.}, import and export rates that are the same for both species. In this scenario, the single stable fixed point (\ref{eq:deterministicmean}) of the deterministic dynamics is located at $y = 0$ (equal concentrations) and $s=1$ (by definition) [see \cref{fig:vecflows} (b)]. 

However, Gillespie simulations of the Master equation at $D\ll1$--- \textit{i.e.}, where the characteristic rate constant of import/export is much less than that of autocatalysis--- are markedly different from such deterministic dynamics in the presence of uncorrelated noise.  Whilst the total population fluctuates around $s=1$ as might have been naively expected, the dynamics of $y$ is characterized by a stochastic switching between $+1$ and $-1$ [see \cref{fig:vecflows} (a)].
This behavior results from the non-trivial state-dependent correlations, and is referred-to as a noise-induced effect.

To capture such behavior, we turn to the Fokker Planck equation that corresponds to (\ref{eq:sSDE}) and (\ref{eq:ySDE}). This is generically written as $\partial_t P(s,y,t) = -\nabla \cdot \bm{J}$, with $\nabla^T=(\partial_s,\partial_y)$ and
\begin{align}
    \bm{J} = \bm{A}P - \frac{1}{2}\nabla^T\cdot(\mathsf{B}P), \label{eq:currentJ}
\end{align} 
the probability current. Numerically approximating this PDE using finite elements \cite{verdugo_software_2022} results in a distribution that is bimodal, such that none of its maxima correspond to the deterministic fixed point [see \cref{fig:vecflows} (c)]. 

Notably, the stationary distribution in $y$ can be found in the joint limit $D\to0$ and $V\to\infty$ such that $\lambda=DV/2$ remains finite.
In this limit, only the first term of $\bm{A}$ and the second term of $\mathsf{B}$ in \eqref{eq:AandB} survive.
The resulting dynamics in $s$ are deterministic, settling at the fixed point $s=1$. Meanwhile, the SDE in the relative proportion becomes
\begin{align}
    \dot{y} &= - y + \sqrt{\frac{1-y^2}{\lambda}}\eta(t) \label{eq:nibistabley}
\end{align}
where $\eta(t)$ is a Gaussian white noise. The steady state solution to this is $P(y)\propto (1-y)^{-1+\lambda}$ which has the classic U-shaped \cite{biancalani_noiseinduced_2012} profile for $\lambda<1$.
The case of $\beta\neq0$ and $\kappa=0$ has been shown to give qualitatively similar results in \cite{bibbona_stationary_2020}, with the distribution in $y$ becoming asymmetric but still bimodal. 

\subsection{Species dependent export}

With the symmetric case understood, we now consider $\kappa \neq 0$--- \textit{i.e.}, export rates that are \textit{not} the same for both species. (Note: for now we keep $\beta = 0$, although asymmetric import is treated later in \cref{sec:asym}). Although nonzero $\kappa$ shifts the location of the deterministic fixed point (\ref{eq:deterministicmean}) it does not change the structure: there is still a single stable node [see \cref{fig:vecflows} (e)]. However, Gillespie simulations at $D\ll1$ again reveal non-trivial behavior. In particular, we see stochastic switching in $y$ that is coupled to periods of sustained growth or decay in $s$ [see \cref{fig:vecflows} (d)].

Again, the Fokker Planck equation proves instructive; whilst the stationary distribution is bimodal, it is no longer symmetric in $y$, with distinct marginals $\int_{-1}^0P(s,y)\dd y$ and $\int_{0}^1P(s,y)\dd y$. Moreover, the current, depicted by streamlines of constant flux [see \cref{fig:vecflows} (f)], now has a clear circulatory structure, indicating a preferred direction of travel for individual trajectories, which perform stochastic loops. Despite similarities with other types of stochastic cycle, such as stochastic amplification and stochastic limit cycles, the behavior we see is aperiodic, with `return' times that are exponentially distributed.

In the following, we describe how to analytically characterize these stochastic loops and the resulting marginal distributions.

\section{\texorpdfstring{$\lambda\ll1$:}{} Exploiting a separation of timescales}

When $\lambda$ is small, we observe stochastic switches in $y$ that are `fast' compared to the dynamic behavior in the total population, $s$, which evolves on a slower timescale (see \cref{fig:vecflows} (d)). 
It is this small $\lambda$ regime that we seek to describe mathematically.
The periods of slow growth and decay are clearly coupled to the fast switches in the relative make-up of the molecules.
Anticipating a separation of timescales, we wish to understand the instantaneous rate at which these switches occur by considering the transition times for $y$ to change from -1 to 1 for a fixed $s$.
By taking the limit $D\to0$ and $V\to\infty$ where $\lambda=DV/2\to 0$, this timescale separation becomes apparent, with the switching time reducing to strongly noise-induced case considered in \cite{biancalani_noiseinduced_2014}. 
In the next section we detail how the switching time can be approximated in this limit and with $\beta=0$, leading to the following switching rate:
\begin{align}
    \gamma(s) = \frac{1}{2s} \label{eq:switchingrate}.
\end{align}
In \cref{sec:PDMP}, this rate is used with a simplified picture of the relative proportion $y$ to characterize the NESS and the transitions between bistable states seen in \cref{fig:vecflows}(e,f).

\subsection{Mean switching time}
\label{sec:switchingtimes}

Directly taking $\lambda\to0$ in (\ref{eq:sSDE},\ref{eq:ySDE}) leads to the same noise strength as in \cref{eq:nibistabley}. As mentioned previously, this leads to a diverging stationary distribution at the boundaries, and thus infinite mean switching time between boundaries.
To ensure a finite switching time, we follow \cite{biancalani_noiseinduced_2014} in keeping a small term, $\varepsilon_D(s)\ll1$, in the noise strength.
It is only once an expression for the switching time is found in terms of $\lambda$ and $\varepsilon_D$ that we take the limit $\lambda\to0$, and by extension $D\to0$.
Meanwhile, taking $V\to\infty$, the first term of the noise correlation $\mathsf{B}$ given in \eqref{eq:AandB} become negligible, as well as the second term of $\bm{A}$. 
Hence, we propose that the SDE for $y$ \eqref{eq:ySDE} can be approximated by 
\begin{align}
    \dot{y} = \mathcal{A}(y) + \sqrt{\frac{\mathcal{B}(y)}{\lambda}}\eta(t), \label{eq:ySDEgeneral}
\end{align}
where
\begin{equation}
\begin{aligned}
    \mathcal{A}(y) &=  -\frac{y}{s}-\kappa(1-y^2), \\
    \mathcal{B}(y) &= 1 + \varepsilon_D - y^2
\end{aligned}
\label{eq:reducedy}
\end{equation}
and $\eta(t)$ is a standard Gaussian white noise.

The additional term in the noise strength can be shown to vanish as $D\to0$ by considering its behavior close to the boundaries.
Away from the boundaries, the noise in $y$ is dominated by the second term in $\mathsf{B}_{22}$ and so $\varepsilon_D$ has little impact.
As $y$ approaches the boundaries, however, the second term in $\mathsf{B}_{22}$ becomes comparable to the first. 
Taking an arbitrary point close to the boundary $|y|=1-Dz$ with $z\leq1$ and Taylor expanding for small $D\ll1$, we compare the full noise strength and the proposed reduced noise strength 
\begin{align*}
    \frac{\mathcal{B}(y)}{\lambda}-\mathsf{B}_{22}(s,y)= \frac{2}{V}\left(\frac{\varepsilon_D}{D} - \frac{1}{s^2}\right) + \mathcal{O}({D}).
\end{align*}
Hence, $\mathcal{B}(y)$ captures the behavior in the noise-dominated case with $\varepsilon_D = D/s^2$. 

To find the rate at which a switch in the dominant species occurs, we find the expected transition time for the system to leave $y=-1$ and reach $y=1$ for a fixed $s$. The inverse of this time, $\gamma(s)$, is the instantaneous rate at which transitions occur.
Treating $y=-1$ as a reflecting boundary and $y=1$ as an absorbing boundary, \cite{gardiner_handbook_1985} gives the expected transition time for a general SDE of the form \cref{eq:ySDEgeneral} as
\begin{align}
    T_\lambda = 2\lambda\int_{-1}^1 \psi(y) \left(\int_{-1}^y \frac{\psi(z)}{\mathcal{B}(z)}\dd z\right) \dd y,
\end{align}
where
\begin{align}
    \psi(y) = \exp(2\lambda\int_{-1}^y \frac{\mathcal{A}(\zeta)}{\mathcal{B}(\zeta)}\dd \zeta). \label{eq:psiy}
\end{align}
Substituting \cref{eq:reducedy} into \cref{eq:psiy}, we find that $\psi(y) = \psi_0(y)\psi_\kappa(y)$ where 
\begin{align*}
    \psi_0(y) =&\; \left(\frac{1+ \varepsilon_D - y^2}{\varepsilon_D }\right)^{ \frac{\lambda}{s}}, \\
    \psi_\kappa(y) = &\; \exp\bigg( 2\kappa \lambda(y+1) \\ 
    &+ \frac{2\kappa\lambda\varepsilon_D}{\sqrt{1+\varepsilon_D}}\arctanh\left(\frac{(1+y)\sqrt{1+\varepsilon_D}}{1+\varepsilon_D+y}\right)\bigg),
\end{align*}
are the symmetric and asymmetric components, respectively.

Now, as $\lambda\to0$, the symmetric component, $\psi_0(y)$, remains finite and non-zero for $y\neq\pm1$ whilst $\lim_{\lambda\to0}\psi_\kappa(y) = 1$. In this limit, therefore, we can calculate the mean switching time from
\begin{align*}
    \lim_{\lambda\to0}T_\lambda = &\; \int_{-1}^1\frac{2\lambda \;\dd y}{\left(1+\varepsilon_D-y^2\right)^{\lambda/s}} \int_{-1}^y \left(1+\varepsilon_D-z^2\right)^{\frac{\lambda}{s}-1}\dd z.
\end{align*}
In summary, for sufficiently small $\lambda$, we need only consider the symmetric components in the switching time.
We note that this reduced integral for the switching time is exactly the same form as in \cite{biancalani_noiseinduced_2014} which considers a similar noise-induced bistable system. 
This integral can be solved for any $\lambda>0,\;\varepsilon_D\ll1$ but, remembering that this approximation only holds for $\lambda\ll1$ and as $\varepsilon_D\to0$, we take its limiting value
\begin{align}
    T_0 = \lim_{\lambda\to0}\frac{2\pi \lambda s}{s-2\lambda}\cot \left(\frac{\pi \lambda}{s}\right) = 2s.
\end{align}
Hence, taking $\gamma(s)=1/T_0$, the rate of switching is given by \cref{eq:switchingrate}.

\begin{figure}[t]
    \centering
    \includegraphics[width=8.6cm]{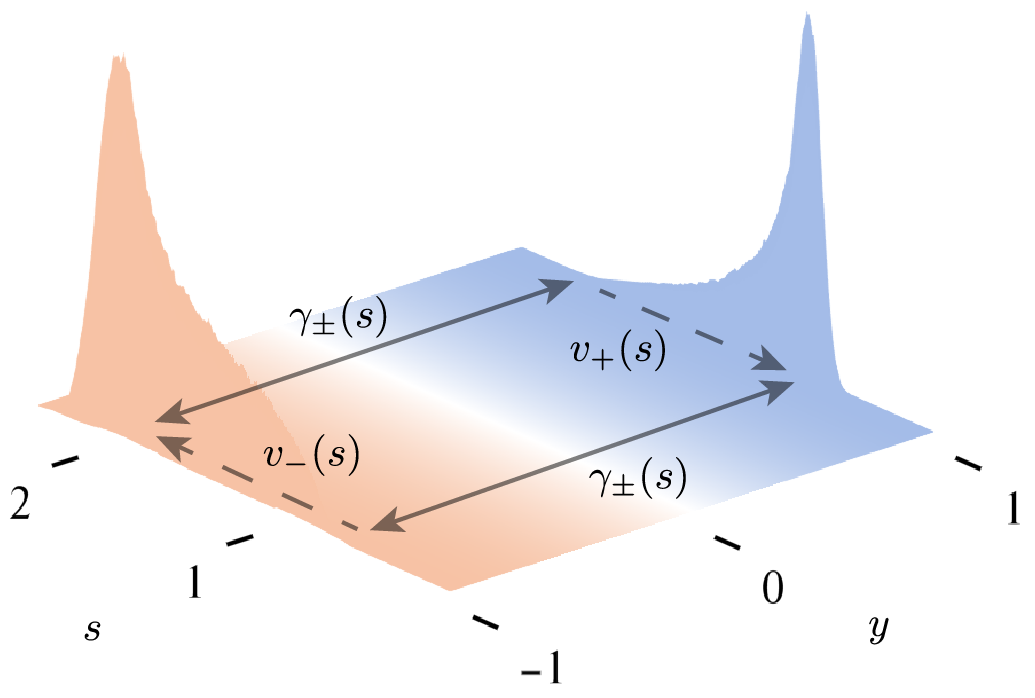}
    \caption{\textbf{Piece-wise deterministic Markov process}. Stationary distribution $P(s,y)$ from Gillespie simulations with $\lambda=0.04$. Dashed arrows indicate the direction of `slow' flow along each axis while solid lines represent `fast' stochastic switches occurring at rate $\gamma(s)$.}
    \label{fig:pdmpcartoon}
\end{figure}

\subsection{Piece-wise Deterministic Markov Process}
\label{sec:PDMP}

\begin{figure*}
    \centering
    \subfigure{\includegraphics[width=8.6cm]{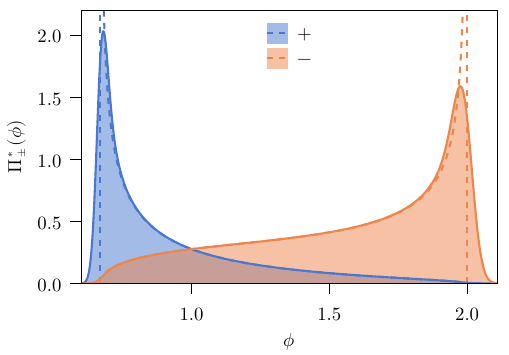}}
    \hfill
    \subfigure{\includegraphics[width=8.6cm]{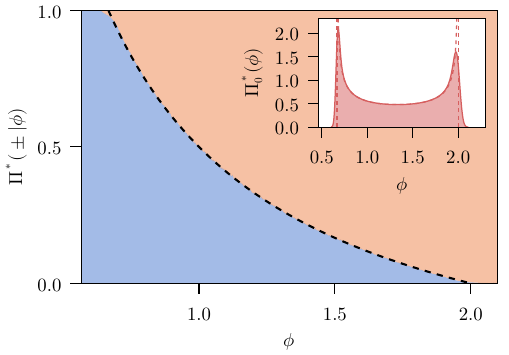}}
    \caption{\textbf{Linear noise approximation}. Analytic stationary solution the PDMP approximation (dashed lines) and numerical approximation to the LNA from \cref{eq:PDMPLNA} (solid lines). The colored regions show the empirical distributions of $x_\pm$ from Gillespie simulations. Left: Stationary distribution for the population in either of the two states. Right: relative probability of the system being dominated by $x_\pm$ for a given $\phi$ with the stationary distribution of $\phi$ in either state in the inset. For both figures, $V=2\times10^3$, with $\lambda=0.01$ and $\kappa=0.5$ and the empirical distributions were determined from 100 independent Gillespie simulations, each with a final (unscaled) time of $t=10^{10}$.}
    \label{fig:PDMP}
\end{figure*}

Here we use the switching behavior discussed in the previous section to inform our analysis of the $s$ dynamics when $0<\kappa<1$.
In this section, therefore, it is assumed that $V$ is sufficiently large that we can neglect finite volume fluctuations in $s$ arising from the noise correlations in \cref{eq:sSDE}.
We also assume that $\lambda\ll1$ such that the prior analysis is applicable here.
Because of this, fluctuations in $y$ either decay rapidly or cause a fast, macroscopic switch in the population proportions. 
At the coarsest level, then, we can think of the system as being in two states, either all $X_+$ or all $X_-$ with Markovian switches between the two. 
We can now begin to describe the dynamics of the total population in terms of these approximate states. 

We denote the total population at this level as $\phi$ to distinguish it from the finite-volume total population, $s$, to give
\begin{align}
    \dot{\phi} = v_{\sigma(t)}(\phi) \equiv 1 - k_{\sigma(t)} \phi .\label{eq:PDMPflow}
\end{align}
where $\sigma(t)\in\{+,-\}$ refers to the state of the relative proportion. 
Because the flow of $\phi(t)$ depends on the Markov process $\sigma(t)$, this is known as a Piecewise Deterministic Markov Process.
A representation of this approximation is displayed in \cref{fig:pdmpcartoon} in the context of the TK2 model.
For $0<\kappa<1$, the flows $v_\pm(\phi)$ have fixed points at $\phi_\pm^*=k_\pm^{-1}$.
By considering the direction of flow in each state, we find that given an initial value, $\phi_0\in\Omega\equiv(\phi_+^*,\phi_-^*)$, $\phi(t)\in\Omega$ for all $t>0$.
Assuming this initial condition, we write $\Pi_\pm(\phi)$ as the probability observing the system with a total population $\phi$ and in a state $\pm$.
All probability distributions resulting from a PDMP are denoted in this way, with the dependence on time suppressed for notational convenience. 
We write the evolution of this probability as a Forward Kolmogorov equation 
\begin{equation}
\begin{aligned}
    \partial_t\Pi_\pm(\phi) =\; &\left[\gamma_-(\phi)\Pi_{\mp}(\phi) - \gamma_+(\phi)\Pi_\pm(\phi)\right] \\ & -\partial_\phi\left[v_\pm(\phi)\Pi_\pm(\phi)\right],
    \label{eq:PDMP}
\end{aligned}
\end{equation}
where, for now, we have left the switching rates between the two states as distinct functions of the population, $\gamma_\pm(\phi)$.
The stationary distribution for this PDMP --- which we generally denote with an asterisk ---  is given in \cite{hufton_intrinsic_2016} as
\begin{align}
    \Pi^*_\pm(\phi) &= \mp C\frac{h(\phi)}{v_{\pm}(\phi)}, \quad \phi\in\Omega.
    \label{eq:PDMPssAnsatz}
\end{align}
The normalization constant, $C$, is defined such that the probability of being in either state, $\Pi_0(\phi)=\Pi_+(\phi)+\Pi_-(\phi)$, is normalized,
\begin{align}
    \int_\Omega\Pi_0(\phi)\dd \phi = 1 .
\end{align}
A full derivation of the stationary distribution is given in \cite{hufton_intrinsic_2016} and \cref{app:PDMPSS}, but it can be verified  simply by substituting \cref{eq:PDMPssAnsatz} into \cref{eq:PDMP} at stationarity. 
We find that it is a solution if
\begin{align*}
    \frac{h'(\phi)}{h(\phi)} = -\left(\frac{\gamma_+(\phi)}{v_+(\phi)}+\frac{\gamma_-(\phi)}{v_-(\phi)}\right).
\end{align*}
The required form of $h(\phi)$ is thus
\begin{align}
    h(\phi) = \exp\left\{-\int^\phi \left(\frac{\gamma_+(u)}{v_+(u)}+\frac{\gamma_-(u)}{v_-(u)}\right)\dd u\right\}. \label{eq:hphi}
\end{align}

Here, we use the symmetric switching rate, $\gamma_\pm(\phi)=\gamma(\phi)=1/2\phi$, found in \cref{sec:switchingtimes}.
Substituting this into \eqref{eq:hphi} along with the expressions for the flows \cref{eq:PDMPflow} we find, after splitting into partial fractions and integrating,
\begin{equation*}
    h(\phi) = \phi^{-1}\sqrt{(k_+\phi-1)(1-k_-\phi)}.
\end{equation*}
The normalization constant can also be determined analytically (see \cref{app:norm}), leading to the full solution:
\begin{align}
    \Pi^*_\pm(\phi) = \frac{\sqrt{1-\kappa^2}}{2\pi\kappa\phi}\left(\frac{k_+\phi-1}{1-k_-\phi}\right)^{\mp\frac{1}{2} }, \quad \phi\in\Omega. \label{eq:PDMPsolSym}
\end{align}
The typical shape of the two distributions is shown in \cref{fig:PDMP} (Left) with the corresponding stationary distributions measured empirically from Gillespie simulations.
The marginal distribution for the total population is given by
\begin{align}
    \Pi_0^*(\phi) = \frac{1}{\pi}\sqrt{\frac{1-\kappa^2}{(k_+\phi-1)(1-k_-\phi)}}. \label{eq:marginalPi0}
\end{align}
The probability of finding the system in a state given a population, $\phi$, is independent of the switching rate:
$\Pi^*(\pm|\phi) =\Pi^*_\pm(\phi)/\Pi^*_0(\phi)$. In this case we have
\begin{align}
    \Pi^*(\pm|\phi) = \frac{v_\mp(\phi)}{2\kappa\phi} =\pm\frac{1-(1\mp\kappa)\phi}{2\kappa\phi}, \label{eq:probStateGivenPhi}
\end{align}
for $\phi\in\Omega$, which is shown in \cref{fig:PDMP} (Right) with the inset axis displaying \cref{eq:marginalPi0}.
If $\kappa=0$, there is no spread in $\phi$ at stationarity as $\phi_+^*=\phi_-^*=1$. Thus, conditioning on $\phi=1$ we see that $\Pi^*(\pm|\phi=1;\kappa=0)=1/2$, which is the symmetric case previously studied in \cite{bibbona_stationary_2020, biancalani_noiseinduced_2012}.

\subsection{Finite volume fluctuations}

In \cref{fig:PDMP}, there are clear deviations from the PDMP solution. These discrepancies are most obvious at the extremities of the stationary distribution, where the total population has a nonzero probability of being outside of the bounds predicted by the PDMP, $\Omega$.
The cause can be traced back to \cref{eq:PDMPflow}, where we assumed the total population was deterministic outside of the macroscopic changes in composition.
In reality, the stochasticity in the exchange of molecules between the cell and reservoir causes the total population to fluctuate, and is the cause of the macroscopic switches, as has been accounted for in \cite{herrerias-azcue_consensus_2019,biancalani_noiseinduced_2012}. 
In other words, we imposed the limit $V\to\infty$ when in reality the effect of finite population sizes is significant.
To account for this, we still consider the system to be in one of the dominant states, but we now readmit the $\mathcal{O}(V^{-\nicefrac{1}{2}})$ fluctuations. 
As $V$ decreases, the fluctuations about the deterministic flow used in \cref{eq:PDMPflow} become more significant. 
We write $\xi=\sqrt{V}(s-\phi)$, and employ a linear noise approximation (LNA) about \cref{eq:sSDE} as has been done for the study of other systems with intrinsic noise subject to fluctuating environments \cite{hufton_intrinsic_2016}. To zeroth order in $V$, we recover the deterministic flow \eqref{eq:PDMPflow} but at the level $\order{V^{-\nicefrac{1}{2}}}$ we find
\begin{align}
    \dot{\xi}(t) = -k_\pm\xi + \sqrt{w_\pm(\phi)}\eta(t),
    \label{eq:LNASDE}
\end{align}
where $w_\pm(\phi)=1+k_\pm\phi$ and $\eta(t)$ is Gaussian white noise.
While this may at first seem like a simple Ornstein-Ulhenbeck process \cite{gardiner_handbook_1985}, we must consider the Markov switching in the state dependent flow and diffusion terms. 
To do so formally would require solving the Fokker Planck equation for the joint distribution of the state, the resulting flow and the fluctuation $P_\pm(\xi,\phi)$. 
Dropping the dependence on the fluctuation and flows for notational simplicity, the Fokker Plank equation is
\begin{equation}
\begin{aligned}
    \partial_t P_\pm = & \; -\partial_\phi \left[v_\pm(\phi)P_\pm\right] + k_\pm\partial_\xi\left[\xi P_\pm\right] \\
    & +\frac{1}{2}\partial^2_\xi\left[w_\pm(\phi) P_\pm\right] + \gamma(\phi)(P_{\mp}-P_{\pm}). \label{eq:fullFokkerPlank}
\end{aligned}
\end{equation}
However, we assume that $P(\xi|\phi,\pm)=P(\xi|\phi)$,
which is true for fluctuations attributed to import events, as these occur at a constant rate. 
This is strictly not the case for export events, on the other hand, since these occur at state-dependent rates and would only hold if $\kappa=0$,
but this source of error is not significant for small $\lambda$.
Practically, this enables us to separate considerations of the fluctuations from the macroscopic state switches by writing $P^*_\pm(\xi,\phi)=P^*(\xi|\phi)\Pi^*(\pm|\phi)\Pi^*_0(\phi)$. Substituting this assumption into \cref{eq:fullFokkerPlank} at stationarity gives
\begin{equation*}
\begin{aligned}
    0 = & \; -P^*(\xi|\phi)\partial_\phi \left[v_\pm(\phi)\Pi^*_0(\phi)\Pi^*(\pm|\phi)\right] \\
    & + k_\pm\Pi^*_0(\phi)\Pi^*(\pm|\phi)\partial_\xi\left[\xi P^*(\xi|\phi)\right] \\
    & +\frac{1}{2}w_\pm(\phi)\Pi^*_0(\phi)\Pi^*(\pm|\phi)\partial^2_\xi P^*(\xi|\phi) \\
    & + \gamma(\phi)\Pi^*_0(\phi)P^*(\xi|\phi)\big(\Pi^*(\mp|\phi)-\Pi^*(\pm|\phi)\big).
\end{aligned}
\end{equation*}
Summing over the two states and substituting \cref{eq:probStateGivenPhi}--- and noticing that $v_+(\phi)\Pi^*(+|\phi)=-v_-(\phi)\Pi^*(-|\phi)$ ensures the $\partial_\phi$ terms cancel--- we have
\begin{equation*}
\begin{aligned}
    0 = & \;  \left(k_+v_-(\phi) - k_-v_+(\phi)\right)\partial_\xi\left[\xi P^*(\xi|\phi)\right] \\
    & +\frac{1}{2}\left(w_+(\phi)v_-(\phi) - w_-(\phi)v_+(\phi)\right)\partial^2_\xi P^*(\xi|\phi). \label{eq:ssFokkerPlankLNA}
\end{aligned}
\end{equation*}
This ordinary differential equation can be solved directly, with $\lim_{\xi\to\pm\infty}P(\xi|\phi)=0$ as boundary conditions. 
Alternatively, we recognize that this is equivalent to the stationary Fokker Planck equation for an Ornstein-Ulhenbeck process and thus the solution is a zero-mean Gaussian distribution with variance given by
\begin{align*}
    \mathrm{Var}(\xi|\phi) = \frac{1}{2}\frac{w_+(\phi)v_-(\phi)-w_-(\phi)v_+(\phi)}{k_+v_-(\phi)-k_-v_+(\phi)}.
\end{align*}
Upon substitution for the flow and noise strength, the variance reduces to $\mathrm{Var}(\xi|\phi)=\phi$.
The total population can be constructed as the probability of the system being in a state $s=\phi+\xi/\sqrt{V}$, which can be calculated by evaluating
\begin{align*}
    P^*_\pm(s)= \iint P(\xi|\phi)\Pi^*_\pm(\phi)\delta\left(\sqrt{V}(s-\phi)-\xi\right)\dd \xi \dd \phi.
\end{align*}
This can be found by numerically integrating
\begin{align}
    P^*_\pm(s) & = \int_\Omega \frac{\Pi^*_\pm(\phi)}{\sqrt{2\pi\phi}}\exp(-\frac{V}{2\phi}(s-\phi)^2)\dd \phi \label{eq:PDMPLNA}.
\end{align}

\subsection{Asymmetric import rates}
\label{sec:asym}

\begin{figure}[t]
    \centering
    \includegraphics[width=8.6cm]{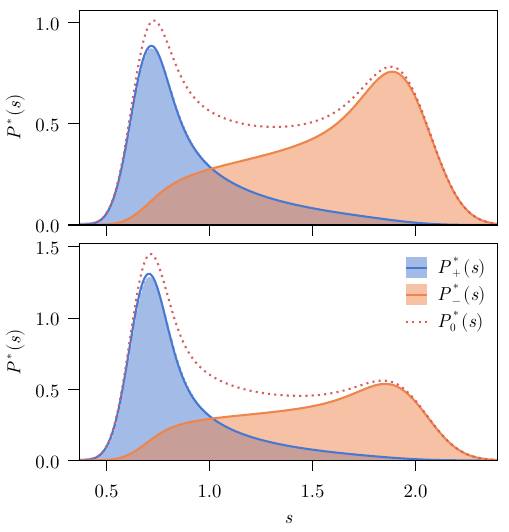}
    \caption{\textbf{Symmetric vs.~asymmetric import}. Linear noise approximation convolved with the PDMP solution given in \cref{eq:PDMPLNA}. Top: symmetric import rates ($\beta=0$). Bottom: asymmetric import rates ($\beta=0.2$) Solid and dotted lines show the numerical solution of \cref{eq:PDMPLNA}. Shaded regions indicate the empirical distributions from 100 independent Gillespie simulations until $t=10^8$. For both figures the asymmetry in export is $\kappa=0.5$ with volume $V=100$ and $\lambda=0.01$.}
    \label{fig:LNA}
\end{figure}

\begin{figure*}
    \centering
    \includegraphics[width=17.2cm]{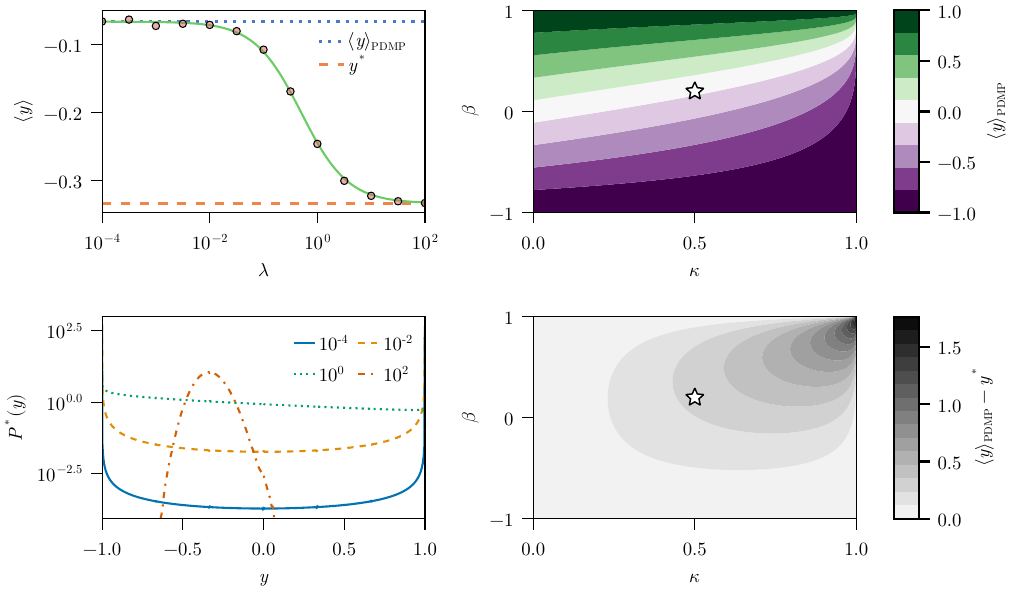}
    \caption{\textbf{Accuracy of the PDMP$+$LNA approximation}. Differences between the deterministic fixed point and the expectation of $y$ in the bistable limit. Top left: expectation of the relative proportion, $y$, as $\lambda$ increases from the noise-induced limit (PDMP approximation) to the deterministic limit, $y^*$; circles indicate stochastic Gillespie simulations and prediction from \cref{eq:transitionPrediction} (solid green). Bottom left: Stationary distributions for $y$ for different $\lambda$ and $V=10^3$. Top right: stationary expectation of $y$ in the PDMP approximation. Bottom right: Difference in the PDMP approximation and the deterministic limit for the expectation of $y$. The stars on the right at  $\beta=0.2$, $\kappa=0.5$ indicate the parameters used for the figures on the left.}
    \label{fig:ycomp}
\end{figure*}

The case of  $\beta\neq0$ has been studied in \cite{bibbona_stationary_2020} but for symmetric $k_\pm$. With the theory discussed above, we can now elucidate the fully asymmetric case by allowing both $\beta\neq0$ and $\kappa>0$. The drift for $y$ in \eqref{eq:ySDEgeneral} is now
\begin{align*}
    \mathcal{A}(y) = \frac{\beta - y}{s} - \kappa(1-y^2),
\end{align*}
and the effect on the noise structure is all on the scale of $\varepsilon_D$ and so we leave $\mathcal{B}(y)$ unchanged. 
However, this minor change still renders the transition time analytically intractable.
To circumvent this problem, we propose a heuristic argument for how the switching rate is altered by this asymmetry.
For models with a fixed system size $N$, exact transition times can be calculated directly from the discrete transition $n_+=0$ to $n_+=N$, described by the Master Equation \cite{houchmandzadeh_exact_2015a,saito_theoretical_2015}.
In this context, when the equivalent $\lambda\to0$ limit is taken, the leading order term found for the transition time arises from the inverse of the expected residence time in the $n_+=0$ ($y=-1$) state. 
The higher order terms relate to the time to transition from the $n_+=1$ to $n_+=N$ states as well as corrections to the residence time in $n_+=0$ due to additional and infrequent spontaneous reactions.
Returning to the general TK2 model, to leading order in $\lambda$, the switching time for the transition $-\to+$ is proportional to the residence time in the $n_+=0$ state. 
Since the residence time is inversely proportional to the rate at which the system leaves $n_+=0$--- and the only way the system leaves such a state is from the import of $X_+$--- it follows that the switching time is proportional to $c_+$. 
Hence, we propose that the instantaneous rates for \emph{leaving} the states $y=\pm1$ are
\begin{align*}
    \gamma_\pm(s) = \frac{c_\mp}{s} = \frac{1\mp\beta}{2s}.
\end{align*}

After substituting the expressions for $\gamma_\pm(\phi)$ and $v_\pm(\phi)$ into \cref{eq:hphi} we find,
\begin{align}
    h(\phi) = \phi^{-1}\left(k_+\phi-1\right)^{(1-\beta)/2}\left(1-k_-\phi\right)^{(1+\beta)/2}, \label{eq:hphiasym}
\end{align}
which leads to the following closed form solution for the PDMP steady state distribution,
\begin{align}
    \Pi^*_\pm(\phi) =\frac{C}{\phi}\left(\frac{k_+\phi-1}{1-k_-\phi}\right)^{\frac{\beta\mp1}2}, \quad \phi\in\Omega. \label{eq:PDMPsolAsym}
\end{align}
In this case, the normalization constant is given explicitly by 
\begin{align*}
    C = \frac{1+\kappa}{2\pi\kappa} \left(\frac{1-\kappa}{1+\kappa}\right)^{\frac{1-\beta}2} \cos\left(\frac{\pi\beta}{2}\right),
\end{align*}
with a full derivation given in \cref{app:norm}.
As the bottom of \cref{fig:LNA} shows, when the LNA approximation is combined with this more general PDMP solution, we can still observe bimodal distributions in $P^*_0(s)=P^*_-(s)+P^*_+(s)$.

\subsection{Average population and relative proportion in the noise-induced limit}

The TK2 system displays drastically different stationary distributions depending on the value of $\lambda$, as both $s$ and $y$ can display bistability when $\lambda$ is sufficiently small but both distributions are Gaussian when $\lambda$ becomes very large.
In this section, we quantify the changes in the expectation values as $\lambda$ changes, which enables us to find the transition region between the noise-induced and deterministic limits.

Initially, we consider the stationary expectation of the total population, $\langle s \rangle$.
We define $\langle\phi\rangle= \int\phi \Pi^*_0(\phi) \dd\phi$ and carry out the integration in a similar manner to  finding the normalization constant of the PDMP (see \cref{app:mean}). 
We find that
\begin{align*}
    \langle\phi\rangle = \frac{1-\beta  \kappa}{1-\kappa ^2},
\end{align*}
matching the deterministic limit \eqref{eq:deterministicmean}. 
We also note that the LNA approximation does not alter this expectation, since the convolution with the Gaussian fluctuations in \cref{eq:PDMPLNA} does not impact the mean and so $\langle s\rangle=s^*$.

The relative proportion, on the other hand, is not invariant with respect to $\lambda$.
In terms of the PDMP approximation, the relative proportion is assumed to be in one of two states, $y=\pm1$, and so its expectation is given by $\langle y\rangle_\textrm{PDMP}= \Phi^*_+ - \Phi^*_-$ where 
\begin{align}
    \Phi^*_\pm = \int_\Omega \Pi^*_\pm(\phi) \dd\phi. \label{eq:probPM}
\end{align}
Again, this can be calculated analytically (see \cref{app:mean}) to give 
\begin{align*}
    \Phi^*_\pm = \mp\frac{1}{2\kappa}\left[1\mp\kappa + (1-\kappa)^{\frac{1-\beta}{2}}(1+\kappa)^{\frac{1+\beta}{2}}\right].
\end{align*}
The difference of these two probabilities is thus 
\begin{align}
    \langle y \rangle_\textrm{PDMP} = \frac{(1-\kappa)^{\frac{1-\beta}{2}}(1+\kappa)^{\frac{1+\beta}{2}}-1}{\kappa} ,\label{eq:expy}
\end{align}
which is displayed in \cref{fig:ycomp} (Top right) as a function of $(\kappa,\beta)$.
It is clear that, in general, $\langle y\rangle_\textrm{PDMP}\neq y^*$ as shown in \cref{fig:ycomp} (Bottom right).
This difference becomes particularly stark for $0.5<\kappa<1$, $0<\beta<1$ as the deterministic limit predicts a positive value but the PDMP limit shows that the system is at $y=+1$ for significant proportion of the time.
As $\kappa\to0$, however, the behavior in $s$ decouples from $y$ and the dynamic bistability reduces to simple noise-induced bistability in $y$ with trivial fluctuations in $s$. 
Without the deterministic flow in \cref{eq:PDMP}, the forward Kolmogorov equation becomes a simple telegraph process with the probability of being in either state given by $\lim_{\kappa\to0}\Phi^*_\pm = (1\pm\beta)/2.$
The expected relative proportion in this case is just $\langle y \rangle_\textrm{PDMP} = \beta$, which is the same as the setting $\kappa=0$ in the deterministic limit \cref{eq:deterministicmean}.

In general, whilst the total population is maintained as $\lambda$ varies, the average relative proportion transitions from the deterministic limit to the noise-induced limit, as can be seen from the simulations depicted in \cref{fig:ycomp} (Top left). 
This demonstrates the ranges for which this system can be characterized, with an intermediate range of $\lambda\lesssim1$ that is neither predominantly noise-driven nor deterministically driven with trivial fluctuations.
In this region, we observe relatively flat distributions in the relative proportion (see the bottom left of \cref{fig:ycomp}), similar to the exactly flat distribution found in the symmetric TK model \cite{biancalani_noiseinduced_2012}.
We are able to find an empirical approximation for $\langle y\rangle$ in this intermediate regime, however. Interpolating between the limiting values: $\lim_{\lambda\to\infty}\langle y \rangle=y^*$ and $\lim_{\lambda\to0}\langle y \rangle= \langle y \rangle_{\textrm{PDMP}}$, we find that
\begin{align}
    \langle y \rangle = \frac{\langle y\rangle_\textrm{PDMP} + 2\lambda y^*}{1+2\lambda}. \label{eq:transitionPrediction}
\end{align}
shows good agreement with Gillespie simulations for a wide range of $\lambda$ values (see \Cref{fig:ycomp} (Top Left)).

\section{Discussion}

By taking the export rates of in the TK2 model to be species-dependent, we have extended its applicability to more biologically relevant scenarios.  Whilst the subject of molecular export and/or secretion is highly complex--- involving both the classical mechanisms that are associated with the endomembrane system as well as neoclassical pathways \cite{cross_delivering_2009,pakdel_exploring_2018,dimou_unconventional_2018}--- the overarching behavior is typified by a remarkably high degree of specificity, such that the cell retains apparently exquisite control over the concentrations of its constituent molecules. The same is true of the lytic pathways that control degradation \cite{groll_molecular_2005}.

In the context of this more general setting, we find that the TK model displays a type of noise-induced growth and/or decay; despite there being only one stable node in the deterministic dynamics, the system stochastically switches between populations comprising almost entirely of one species, whence the total population size either increases or decreases steadily dependent on the species at hand. 
This behavior is shown to be captured by a NESS of the requisite Fokker-Planck equation, in which there is a circulatory current, in contrast with the typical noise-induced metastability observed in systems with fixed, fluctuating, or trivially growing population size.

In our analysis, we exploited a separation of timescales, captured by the limit $\lambda\to0$, to arrive at a simple expression for the instantaneous rate of switching. 
We note, however, that this provides a useful approximation up to $\lambda\lesssim0.1$, with the stationary distribution calculated from the PDMP and LNA convolution matching simulations with good accuracy. 
At or above this point, the marginal probability distribution of $y$ becomes too diffuse to consider the system in one of two macrostates. 
A more complete understanding of the weakly bistable regime $0.1<\lambda<1$ would be of interest but presents a great deal of mathematical complexity.

Although phase space trajectories perform stochastic loops, these are fundamentally different to other types of stochastic cycle, such as those due to stochastic limit cycles or arising from stochastic amplification \cite{boland_how_2008}.
In those cases, the cycles are contingent on oscillations (either persistent, or damped, respectively) in the underlying deterministic dynamics.
Here, the deterministic dynamics has no such features, and there is therefore no peak in the power spectrum of stochastic trajectories. In the $\lambda\to 0$ limit, return times are instead exponentially distributed, reflecting the underlying role of Poissonian stochastic switches.

In terms of methods, whilst examples of noise-induced multi-stability in biological settings have been attributed stochastic gene expression \cite{gander_stochastic_2007, raj_stochastic_2008, visco_switching_2010, dubnau_bistability_2006} as well as second order mass-action export \cite{sardanyes_noiseinduced_2018}, PDMPs have been only been employed effectively to model the former \cite{faggionato_nonequilibrium_2009,zeiser_autocatalytic_2010, hufton_model_2019}. As far as the authors of this article are aware, until now they have not been used to describe switches from intrinsic autocatalytic reactions.

Moreover, since the original mathematical treatment of the TK model has been applied to various other systems that display noise-induced metastability \cite{biancalani_noiseinduced_2014,herrerias-azcue_consensus_2019,jafarpour_noiseinduced_2015,jafarpour_noiseinduced_2017,jhawar_noiseinduced_2020}, we anticipate that this work may provide a similar launching point for the modification and subsequent analysis of a variety of models--- typically referred-to as the Voter class--- from the fields of opinion dynamics and collective behavior. Indeed, as our methodology is not limited to studying variable population sizes, and any system that contains secondary state variables which depend on the noise-induced metastable variable could be subjected to the same analysis presented here. We therefore welcome further work in the area.

\section*{Declarations}
The authors have no competing interests to declare.

\begin{acknowledgments}
JRW and RGM acknowledge funding from the EMBL Australia program. RGM acknowledges funding from the Australian Research Council Centre of Excellence for Mathematical Analysis of Cellular Systems (CE230100001).
\end{acknowledgments}

\appendix

\section{PDMP Stationary distribution derivation}
\label{app:PDMPSS}
Here we derive the stationary distribution to the Forward Kolmogorov equation given in \eqref{eq:PDMP} for the PDMP.
A similar derivation of this can be found in \cite{hufton_intrinsic_2016} and an alternative, more rigorous analysis is given in \cite{faggionato_nonequilibrium_2009}. First, we assume each flow has a fixed point $v_\pm(\phi_\pm^*)=0$, and for convenience choose $0<\phi^*_+<\phi^*_-<\infty$ and that $v_+(\phi)<0<v_-(\phi),\;\forall\phi\in\Omega=(\phi^*_+,\phi^*_-)$.
From this, we find zero-current boundary conditions:
\begin{align}
    v_\pm(\phi^*_\pm)\Pi_\pm(\phi^*_\pm) = 0. \label{eq:zerocurrent}
\end{align}
Now, taking \cref{eq:PDMP} at stationarity and summing over the two states we have
\begin{align*}
    \partial_\phi\left[v_+(\phi)\Pi^*_+(\phi)+v_-(\phi)\Pi^*_-(\phi)\right] = 0.
\end{align*}
Hence, integrating from the left boundary and using \eqref{eq:zerocurrent}, this gives
\begin{align*}
    0 &= \int_{\phi_+^*}^\phi\partial_\phi\left[v_+(u)\Pi^*_+(u)+v_-(u)\Pi^*_-(u)\right] \dd u \\
    & = v_+(\phi)\Pi^*_+(\phi)+v_-(\phi)\Pi^*_-(\phi),
\end{align*}
and thus
\begin{align}
    \Pi^*_-(\phi) = -\frac{v_+(\phi)}{v_-(\phi)}\Pi^*_+(\phi). \label{eq:pmrelation}
\end{align}
Next, we substitute this into the $+$ state of \cref{eq:PDMP} at stationarity,
\begin{align*}
    \partial_\phi[v_+(\phi)\Pi^*_+(\phi)] =v_+(\phi)\Pi_+^*(\phi)\left(\frac{\gamma_-(\phi)}{v_-(\phi)} + \frac{\gamma_+(\phi)}{v_+(\phi)}\right).
\end{align*}
The solution to this ordinary differential equation becomes clearer by writing
\begin{align*}
    h'(\phi) =-h(\phi)\left(\frac{\gamma_-(\phi)}{v_-(\phi)} + \frac{\gamma_+(\phi)}{v_+(\phi)}\right),
\end{align*}
where we have introduced
\begin{align}
    h(\phi)=\frac{v_+(\phi)\Pi^*_+(\phi)}{C}, \label{eq:hphiAnzatz}
\end{align}
and $C$ is a normalization constant whose significance and exact form will become clear subsequently.
This can be solved by separation of variables which gives us
\begin{align}
    h(\phi) = \exp(-\int^\phi \left(\frac{\gamma_-(u)}{v_-(u)} + \frac{\gamma_+(u)}{v_+(u)}\right)\dd u). \label{eq:hphiApp}
\end{align}
Hence, the stationary distribution for either state can be summarized by combining \cref{eq:hphiAnzatz} and \eqref{eq:pmrelation} to give
\begin{align*}
    \Pi^*_\pm(\phi) = \mp C\frac{h(\phi)}{v_\pm(\phi)}.
\end{align*}

Since the stationary distribution is a probability density, we require it to be normalized over both $\phi$ and the states $\pm$, which means that
\begin{align*}
    \int_\Omega\Pi_+(\phi)+\Pi_-(\phi)\dd \phi = 1,
\end{align*}
and so 
\begin{align}
    C = \left[\int_\Omega h(\phi)\left(\frac{1}{v_-(\phi)}-\frac{1}{v_+(\phi)} \right) \dd \phi\right]^{-1}. \label{eq:Cconst}
\end{align}

\section{Normalization constant for the PDMP}
\label{app:norm}

Substituting the explicit form of $h(\phi)$ from \cref{eq:hphiasym}, into \cref{eq:Cconst}, we have that
\begin{align}
    C^{-1} = 2\kappa\int_\Omega\left(k_+\phi-1\right)^{-\frac{1+\beta}{2}}\left(1-k_-\phi\right)^{\frac{\beta-1}{2}} \dd\phi, \label{eq:intConst}
\end{align}
where $\phi_\pm^*=1/k_\pm$.
To solve this integral, we first transform to the unit interval by introducing
\begin{align}
    x = \frac{\phi-\phi_+^*}{\phi_-^*-\phi_+^*} = \frac{1-\kappa}{2\kappa}\left[(1+\kappa)\phi - 1 \right]. \label{eq:transformation}
\end{align}
Rearranging for $\phi$, we have
\begin{align*}
    \phi= \frac{1}{1+\kappa}\left(1+\frac{2\kappa x}{1-\kappa}\right).
\end{align*}
Substituting these into \cref{eq:intConst} and after some simplification, we have
\begin{align*}
    C^{-1} & = 2\kappa \int_0^1 \left( \frac{2\kappa x}{1-\kappa} \right)^{-\frac{1+\beta}2} \left( \frac{2\kappa }{1+\kappa}(1-x) \right)^{\frac{\beta-1}{2}} \\
    &\qquad\qquad \times\frac{2\kappa}{(1-\kappa)(1+\kappa)} \dd x.
\end{align*}
Upon further rearrangement, it can be shown that the normalization constant is given by
\begin{align}
     C^{-1} = \frac{2\kappa}{1+\kappa} \left(\frac{1+\kappa}{1-\kappa}\right)^{(1-\beta)/2}  \mathcal{I}_0(\beta),
     \label{eq:CconstApp}
\end{align}
where the integral,
\begin{align}
    \mathcal{I}_0(\beta) = \int_0^1 x^{-(\beta +1)/2} (1-x)^{(\beta -1)/2} \dd x,
\end{align}
remains to be solved for $-1<\beta<1$. 
We compare this to the integral form for Euler's Beta function which, in terms of the Gamma function, is
\begin{align}
    \int_0^1x^{z_1-1}(1-x)^{z_2-1}\dd x = \frac{\Gamma(z_1)\Gamma(z_2)}{\Gamma(z_1+z_2)}, \label{eq:betafunc}
\end{align}
where we identify $z_1=(1-\beta)/2$, $z_2=(1+\beta)/2$, and $\Gamma(z_1+z_2)=\Gamma(1)=1$.
We also use the identity $\Gamma(z)\Gamma(1-z)= \pi/\sin(\pi z)$ along with standard trigonometric rules to obtain
\begin{align*}
    \mathcal{I}_0(\beta) = \pi\sec\left(\frac{\pi\beta}{2}\right).
\end{align*}
Hence, substituting this back into \cref{eq:CconstApp}, the closed form expression for the normalization constant is
\begin{align*}
    C = \frac{1+\kappa}{2\pi\kappa} \left(\frac{1-\kappa}{1+\kappa}\right)^{(1-\beta)/2} \cos\left(\frac{\pi\beta}{2}\right). 
\end{align*}

\section{Mean of the PDMP stationary distribution}
\label{app:mean}

Here we show that the mean population size calculated from the stationary PDMP solution matches the expectation of the van Kampen expansion SDEs \eqref{eq:sSDE}. We consider the two contributions to the mean of $\phi$, from the two macrostates by writing $\mu_0^{(\phi)} = \mu_+^{(\phi)}+\mu_-^{(\phi)}$ where
\begin{align*}
    \mu_\pm^{(\phi)} & = \int_\Omega\phi\Pi_\pm^*(\phi)\dd\phi.
\end{align*}
It is important to note that the contributions themselves are not a measurement of the mean of each state since the integral is implicitly dependent on the probability of being in such a state: $\Pi_\pm^*(\phi)=\Pi^*(\pm)\Pi^*(\phi|\pm)$.
Performing the same transformation as in \Cref{app:norm} using \cref{eq:transformation}, we find that
\begin{align*}
    \mu_\pm^{(\phi)} = \frac{2\kappa C}{1-\kappa^2}\left(\frac{1-\kappa}{1+\kappa}\right)^{\frac{\beta\pm1}2}\mathcal{I}_\pm(\beta),
\end{align*}
where we have introduced
\begin{align*}
    \mathcal{I}_\pm(\beta) = \int_0^1 x^{-(\beta \pm1)/2} (1-x)^{(\beta \pm1)/2} \dd x.
\end{align*}
Substituting the expression for the normalization constant \eqref{eq:CconstApp} and simplifying, the contributions to the mean are
\begin{align}
    \mu_\pm^{(\phi)} = \frac{1}{1\pm\kappa}\frac{\mathcal{I}_\pm(\beta)}{\mathcal{I}_0(\beta)}. \label{eq:meanInt}
\end{align}

Similarly to \Cref{app:norm}, we compare to \cref{eq:betafunc} in for both states $\pm$.
For $I_+(\beta)$, we identify $z_1=(1-\beta)/2,\;z_2=(3+\beta)/2$ and for $I_-(\beta)$ we have $z_1=(3-\beta)/2,\;z_2=(1+\beta)/2$. As a result, we obtain
\begin{align*}
    \mathcal{I}_\pm(\beta) & = \left(\frac{1\pm\beta}{2}\right)\Gamma\left(\frac{1\mp\beta}{2}\right)\Gamma\left(\frac{1\pm\beta}{2}\right) \\
    & = \left(\frac{1\pm\beta}{2}\right)\mathcal{I}_0(\beta).
\end{align*}
where we have used the fact that $\Gamma(z)=(z-1)\Gamma(z-1)$.
Substituting this result into \cref{eq:meanInt} we have
\begin{align*}
    \mu_\pm^{(\phi)} = \frac{1\pm\beta}{2(1\pm\kappa)}.
\end{align*}
The overall mean, is thus the sum of these two contributions
\begin{align*}
    \mu_0^{(\phi)} = \frac{1-\beta\kappa}{1-\kappa^2}.
\end{align*}

\section{Mean relative proportion}

To find the average value of of the relative proportion in the PDMP limit, $\langle y \rangle_\textrm{PDMP}$, we first calculate the probability of being in the + state at stationarity, $\Phi^*_+$, defined in \cref{eq:probPM}. The expectation is then given by $\langle y \rangle_\textrm{PDMP}=\Phi^*_+-\Phi^*_-=2\Phi^*_+-1$. Using the PDMP solution \eqref{eq:PDMPsolAsym} and using the same substitution \eqref{eq:transformation}, we have
\begin{align}
    \Phi^*_+ = \frac{2\kappa C}{1+\kappa} \left(\frac{1+\kappa}{1-\kappa}\right)^{(1-\beta)/2}\mathcal{I}'(\beta,\kappa),\label{eq:pPMApp}
\end{align}
where
\begin{align*}
    \mathcal{I}'(\beta,\kappa) = \int_0^1 x^{-\frac{1+\beta}{2}}(1-x)^{\frac{1+\beta}{2}}\left(1+\frac{2\kappa x}{1-\kappa}\right)^{-1}\dd x.
\end{align*}

This integral is also in the form of Euler's integral formula for the Beta function, but in this case we have
\begin{align*}
    \mathcal{I}'(\beta,\kappa) = \left(\frac{1+\beta}{2}\right)\mathcal{I}_0(\beta) \prescript{}{2}{F}_1\left(1, \frac{1-\beta}{2}, 2; \frac{2\kappa}{\kappa-1}\right),
\end{align*}
where $\prescript{}{2}{F}_1$ is the hypergeometric function. Substituting this, along with \cref{eq:CconstApp}, into \cref{eq:pPMApp}, the expression for the probability reduces to
\begin{align*}
    \Phi^*_+ = \left(\frac{1+\beta}{2}\right)\prescript{}{2}{F}_1\left(1, \frac{1-\beta}{2}, 2; \frac{2\kappa}{\kappa-1}\right).
\end{align*}
This can be written explicitly by recognizing that
\begin{align*}
    \prescript{}{2}{F}_1(1,b,2;z) = \frac{(1-z)^{-b}\left((1-z)^b+z-1\right)}{(b-1)z}.
\end{align*}
Hence, we arrive at
\begin{align*}
    \Phi^*_+ = -\frac{1}{2\kappa}\left[1-\kappa + (1-\kappa)^{\frac{1-\beta}{2}}(1+\kappa)^{\frac{1+\beta}{2}}\right],
\end{align*}
from which the relative proportion is readily obtained:
\begin{align*}
    \langle y \rangle_\textrm{PDMP} = \frac{(1-\kappa)^{\frac{1-\beta}{2}}(1+\kappa)^{\frac{1+\beta}{2}}-1}{\kappa}.
\end{align*}

\end{document}